\documentstyle[aps,pre,multicol,epsf,float]{revtex}
\begin{document}
\title{Roughening and preroughening transitions in crystal surfaces with
double-height steps}
\author{Jae Dong Noh
\thanks{Present Address: Center for Theoretical Physics, 
Seoul National University, Seoul 151-742, Korea}}
\address{Department of Physics, University of Washington, 
Seattle, WA 98195-1560, U.S.A.}
\maketitle
\begin{abstract}
We investigate phase transitions in a solid-on-solid model where 
double-height steps as well as single-height steps are allowed. 
Without the double-height steps, repulsive interactions between 
up-up or down-down step pairs give rise to a disordered flat phase. 
When the double-height steps are allowed, two single-height steps can 
merge into a double-height step (step doubling).
We find that the step doubling reduces repulsive interaction strength 
between single-height steps and that the disordered flat 
phase is suppressed. 
As a control parameter a step doubling energy is introduced, 
which is assigned to each step doubling vertex. 
From transfer matrix type finite-size-scaling studies of 
interface free energies, we obtain the phase diagram in the parameter space of 
the step energy, the interaction energy, and the step doubling energy. 
\end{abstract}

\begin{multicols}{2}
Much attention has been paid to the phase transitions in 
crystal surfaces since they show rich critical phenomena. 
The interplay between roughening and reconstruction
results in interesting phases, such as a disordered flat~(DOF) phase, 
as well as flat and rough phases~\cite{denNijs94}. 
In the DOF phase the surface is filled with macroscopic amount of steps
which are disordered positionally but have up-down order.
Several solid-on-solid~(SOS) type models have been studied,
which reveals that the DOF phase is stabilized by the repulsive step-step 
interactions~\cite{Rommelse&denNijs87_89,Prestipino95,Bastiaansen96} or by 
specific topological properties of surfaces, e.g., Si(001)~\cite{denNijs97}. 

The SOS type model studies have been done in cases where the 
nearest-neighbor~(NN) height difference, $\Delta h$, is restricted to 
be equal to or less than 1 in units of the lattice constant. 
However, in real crystals there also appear steps with $|\Delta h| > 1$.
For example, double-height steps on W(430) become more favorable than 
single-height steps at high temperatures since they have lower kink
energy~\cite{Dey96}. 
In this paper we investigate the phase transitions in crystal surfaces in
the presence of the double-height steps with $|\Delta h| =2$, especially 
focusing on the stability of the DOF phase. 
We study a generalized version of the restricted solid-on-solid~(RSOS) model 
on a square lattice with the Hamiltonian given in Eq.~(\ref{H_rsos5}). 
We study the model under the periodic and anti-periodic boundary conditions, 
from which various interface free energies are defined. 
The interface free energy is calculated from numerical diagonalizations of 
the transfer matrix, and the phase diagram is obtained by analyzing their 
finite-size-scaling~(FSS) properties.

In the RSOS model the surface is described by integer-valued heights 
$h_{\bf r}$ at each site ${\bf r}=(n,m)$ on a square 
lattice.~(The lattice constant in the $z$ direction is set to 1.) 
Only the single-height step~(S step) with $|\Delta h|=1$ is allowed.
It was found that the RSOS model with NN and next-nearest-neighbor~(NNN) 
interactions between height displays the DOF phase 
when the NNN coupling strength is large enough~\cite{Rommelse&denNijs87_89}.
The NNN coupling accounts for the repulsive interactions between
parallel~(up-up or down-down) step pairs. 
Parallel step pairs cost more energy than 
anti-parallel~(up-down or down-up) step pairs. 

\begin{figure}[t]
\centerline{\epsfxsize=8cm \epsfbox{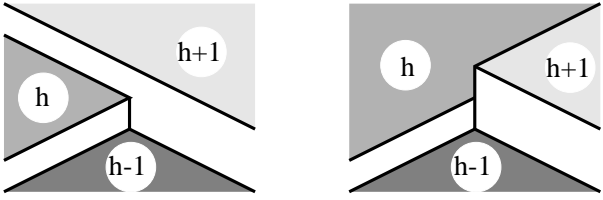}}
\caption{\narrowtext Step doubling vertices.}\label{fig:1}
\end{figure}

The double-height step~(D step) is incorporated into the RSOS model 
by relaxing the restriction on the NN height difference to 
$|\Delta h| = 0, 1, 2$. We only consider quadratic NN and NNN 
interactions between heights since they are sufficient to describe 
the key feature of the phase transitions. The total Hamiltonian is written as
\begin{equation}
H_0 = K \sum_{\langle{\bf r},{\bf r'}\rangle} (h_{\bf r}-h_{\bf r'})^2 +
    L \sum_{({\bf r},{\bf r''})} (h_{\bf r}-h_{\bf r''})^2
\end{equation}
where $\langle \rangle$ and $()$ denote the pair of NN and NNN sites.
With this Hamiltonian, a D step costs more energy than two separate S steps
by an amount of $2K+4K$ per unit length. Even though the D steps are
energetically unfavorable, we will show that their effect is not 
negligible.
We also consider a step-doubling energy $E_D$ to study the effect of 
the step doubling. It is assigned to each vertex where two S steps 
merge into a D step~(see Fig.~\ref{fig:1}). 
The electronic state at step edges may be different from
that at a flat surface, which contributes to the step energy. 
When two S steps merge into a D step, the electronic state near the
vertex may be changed. The change leads to an additional energy cost, which
is reflected by $E_D$.  When $E_D$ is positive~(negative), 
it suppresses~(enhances) the step doubling. 
The Hamiltonian including $H_0$ and the step-doubling energy is then given by
\begin{equation}\label{H_rsos5}
H = H_0 + E_D N_D
\end{equation}
where $N_D$ is the total number of step-doubling 
vertices.~(For a notational convenience the energy is measured in unit 
of $k_BT$.) 
The model with the Hamiltonian Eq.~(\ref{H_rsos5}) with 
$E_D=0$ and with the restriction $|\Delta h| = 0, 1$ will be referred to as 
the RSOS3 model, and the model with the Hamiltonian Eq.~(\ref{H_rsos5}) 
and with $|\Delta h|=0, 1, 2$ will be referred to as the RSOS5 model.

In a continuum description phase transitions in crystal 
surfaces are described by the sine-Gordon model
\begin{equation}\label{H_cont}
H = \int d^2{\bf r} \left[ \frac{1}{2} K_G (\nabla \phi)^2 - 
\sum_{q=1}^\infty u_{q} \cos(2q\pi\phi) \right] \ ,
\end{equation}
where $\phi({\bf r}) \in (-\infty,\infty)$ is a real-valued local average 
height field, $K_G$ the stiffness constant, and $u_{q}$ the fugacity of 
$q$-charge~\cite{denNijs90}. 
In the renormalization group sense $u_1$ is irrelevant at high temperatures 
where the model renormalizes to the Gaussian model with a renormalized
stiffness constant $K_G<\frac{\pi}{2}$ describing the rough phase.  
As temperature decreases, $u_1$ becomes relevant at a roughening transition
temperature. There appear two kinds of low temperature phases depending 
on the sign of $u_1$: 
For positive $u_1$ the Hamiltonian favors an integer average height 
and hence the surface is flat. For a negative $u_1$ it favors a half-integer 
average height. Since the microscopic height is integer-valued, the surface 
can take the half-integer average height by forming steps with up-down order, 
i.e., the surface is in the DOF phase. As temperature decreases further,
the sign of $u_1$ changes and the surface falls into the flat phase.
At the roughening transition between the rough phase and the flat or DOF
phase, the renormalized stiffness constant takes the universal value 
of $\frac{\pi}{2}$. The flat and DOF phases are separated by the preroughening 
transition characterized by $u_1=0$~\cite{denNijs90}.

The phase boundaries can be obtained using FSS properties of the interface 
free energies. Consider the model on a finite 
$N\times M$ square lattice rotated by $45^\circ$ under the various 
boundary conditions~(BC's): The periodic BC, $h(n+N,m) = h(n,m)+a$ with 
integer $a$, and the anti-periodic BC, $h(n+N,m) = -h(n,m)+a \ 
(\mbox{mod\ }2)$ with $a=0 \mbox{ and } 1$.
They will be referred to as $(\pm,a)$ BC's~(the upper~(lower) sign for 
the (anti-)periodic BC's). The free energy is obtained 
from the largest eigenvalue of the transfer matrix.
Detailed description of the transfer matrix set-up can be found in 
Ref.~\cite{Rommelse&denNijs87_89,denNijs97}.
The boundary conditions except for the $(+,0)$ BC 
induce a frustration in the surface. The interface free energy 
$\eta_\kappa$ is defined as the excess free energy per unit 
length under the $\kappa$ BC with $\kappa=(\pm,a)$ 
from that under the $(+,0)$ BC:
\begin{equation}
\eta_\kappa = -\frac{1}{M} \ln \frac{Z_\kappa}{Z_{(+,0)}}
\end{equation}
with $Z_\kappa$ the partition function satisfying the $\kappa$-BC.

The interface free energies have characteristic FSS properties in each
phase. In the rough phase they show the universal $1/N$ scaling in the 
semi-infinite limit $M\rightarrow\infty$ as
\begin{eqnarray}
\eta_{(+,a)} &=& \frac{\zeta}{2} \frac{K_Ga^2}{N} + o\left(
\frac{1}{N}\right) \label{fss_pb} \\
\eta_{(-,a)} &=& \frac{\pi\zeta}{4N} + o\left( \frac{1}{N}\right)  \ ,
\label{fss_apb}
\end{eqnarray}
where $K_G\leq\frac{\pi}{2}$ is the renormalized stiffness constant of 
the Gaussian 
model and $\zeta$ is the aspect ratio of the lattice constants in the 
horizontal and vertical directions~\cite{Rommelse&denNijs87_89,Knops94}. 
In the flat phase $\eta_{(+,a)}$ and $\eta_{(-,1)}$ are finite
because at least one step is induced under the $(+,a)$ and 
$(-,1)$ BC's, while $\eta_{(-,0)}$ is exponentially small in $N$ since 
the $(-,0)$ BC may not induce any steps~\cite{Rommelse&denNijs87_89}. 
In the DOF phase the $(-,1)$ BC does not induce any frustration in the 
step up-down order, but the $(+,a)$ and $(-,0)$ BC's do. 
So $\eta_{(-,1)}$ is exponentially small in $N$, and $\eta_{(+,a)}$ and 
$\eta_{(-,0)}$ are finite~\cite{Rommelse&denNijs87_89}. 
From these FSS properties the roughening points can be estimated from 
\begin{equation}\label{fss_pb_r}
\eta_{(+,1)} = \frac{\pi \zeta}{4N} , 
\end{equation}
where the universal value of $K_G=\frac{\pi}{2}$ at the roughening
transition is used in Eq.~(\ref{fss_pb}). The preroughening 
points between the flat and the DOF phase can be estimated from the crossing
behaviors of $N\eta_{(-,0)}$ or $N \eta_{(-,1)}$, which converges to zero
in one phase and diverges to infinity in the other phase as $N$ grows. 

The estimation of transition points using the interface free energies 
suffers from slow convergence due to corrections to the scaling. 
They may smooth out the crossing 
behaviors of $N\eta_{(-,0)}$ and $N \eta_{(-,1)}$ at the preroughening 
transitions for small $N$. 
But one can safely cancel out leading corrections to scaling by 
taking the ratio or the difference of them, which can be seen as follows. 
Consider the lattice version of the continuum model in Eq.~(\ref{H_cont}).
It is obvious, using the transformation $\phi \rightarrow \phi -1/2$, that 
the model under the $(-,0)$ BC is the same as that under the $(-,1)$ 
BC with $u_{q}$ replaced by $-u_{q}$ for odd $q$. It yields the relation
\begin{equation}
\eta_{(-,0)}(u_1,u_2,u_3,\ldots) = \eta_{(-,1)}(-u_1,u_2,-u_3,\ldots) \ .
\end{equation} 
So if one neglects all higher order contributions from $u_{q\geq 3}$,
the location of $u_1=0$ is found from the condition 
$\eta_{(-,0)}-\eta_{(-,1)}=0$ or $R=1$ with
\begin{equation}\label{R_def}
R \equiv \frac{\eta_{(-,0)}}{\eta_{(-,1)}} \ .
\end{equation} 
It is not influenced by correction-to-scalings from $u_2$. Therefore the
relation $R=1$ can be used to get the $u_1=0$ point more accurately. 
One can easily see that $R>1$ for negative $u_1$ and $R<1$ for positive
$u_1$. It approaches 1 in the rough phase and at the preroughening
transition points, diverges in the DOF phase, and vanishes in the flat
phase as $N\rightarrow\infty$.

\begin{figure}
\centerline{\epsfxsize=70mm \epsfbox{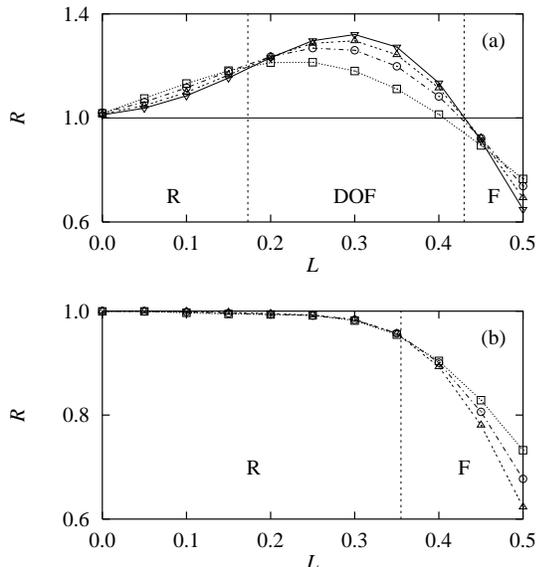}}
\caption{The ratio $R$ defined in the text for the RSOS3 model 
(a) and the RSOS5 model (b) along the line $L=5K$. The rough, DOF, 
and flat phases are denoted by R, DOF, and F, respectively. 
($\Box$ for $N=4$, $\circ$ for $N=6$, $\bigtriangleup$ for $N=8$, and 
$\bigtriangledown$ for $N=10$.)}\label{fig:2}
\end{figure}
In the RSOS3 model the exact point with $u_1=0$ is known along the line 
$L=0$~\cite{denNijs90}.  It is called the self-dual point and is 
located at $K=K_{SD} = \ln [\frac{1}{2}(\sqrt{5}+1)]$. 
From numerical studies of the RSOS3 model transfer matrix, we could 
obtain the exact value of $K_{SD}$ with error less than $10^{-12}$ by 
solving $R=1$ even with small system size $N=4$, which indicates that $R$ is a
useful quantity to determine the preroughening transition points accurately. 
It will be used in the analysis of the RSOS5 model.

We first consider the RSOS5 model in a special case of $E_D=0$ and compare
its phase diagram with that of the RSOS3 model to have insight into 
the role of the D step. At low temperatures
the D step is unfavorable due to larger free energy cost than the 
S step. So the nature of the low temperature phase in the RSOS5 model 
is not different from that in the RSOS3 model, i.e., the flat phase. 
At high temperatures, the surface is in the rough
phase in the RSOS3 model. Since the rough phase is critical and there is no 
characteristic length scale, there will be no difference between S and 
D steps. So the RSOS5 model will also have the rough phase as a 
high temperature phase. 
There is significant difference in intermediate temperature range, where the 
repulsive step interactions stabilize the DOF phase in the RSOS3 model.
Without the D steps the parallel steps have less meandering entropy than 
anti-parallel ones. It is energetically unfavorable for parallel steps to 
approach each other closer than the interaction range while anti-parallel 
steps can approach each other at will~\cite{Rommelse&denNijs87_89}. 
However, if one allows the D step, two parallel S steps can approach each 
other and form a D step without the interaction energy cost. 
Provided that the energy cost of the D step is not too high, the presence 
of the D step reduces repulsive interaction strength effectively and 
enhances the meandering entropy of parallel steps. Then it will suppress 
the DOF phase.

To see such effects quantitatively, we calculate the ratio 
$R$ for the RSOS3 model and the RSOS5 model with $E_D=0$ 
along a line $L=5K$~(see Fig.~\ref{fig:2}). 
The strip width for the transfer matrix is $N=4, 6, 8$,
and $10$ for the RSOS3 model and $N=4,6$, and $8$ for the RSOS5 model. 
The RSOS3 model displays the roughening and the preroughening transitions 
along the line $L=5K$, which is manifest in Fig.~\ref{fig:2}(a). 
There are three regions where $N$ dependence of $R$ 
is distinct with each other. The surface is in the rough phase with 
negative $u_1$ in the small $L$~(high temperature) region, where $R$ 
approaches $1$ from above.
And the surface is in the DOF~(flat) phase for the intermediate~(large) $L$ 
region, where $R$ grows~(vanishes). 
The roughening and preroughening transition points are estimated from
Eq.~(\ref{fss_pb_r}) and $R=1$ with $R$ in Eq.~(\ref{R_def}), respectively, 
which are represented by broken vertical lines.

The situation changes qualitatively in the RSOS5
model. As can be seen in Fig.~\ref{fig:2}(b), $R$ is always less than 1, and 
there are only two regions with distinct $N$ dependence of $R$.
In the small $L$ region $R$ approach $1$ from below, and 
in the large $L$ region $R$ vanishes as $N$ increases. 
They correspond to the rough phase 
with positive $u_1$ and the flat phase, respectively. The roughening
transition point is estimated from Eq.~(\ref{fss_pb_r}) and
represented by the broken vertical line.
It shows that the DOF phase is suppressed in the presence of the 
D step. 
We have also checked that $R$ is always less than 1~($u_1>0$) and 
the DOF phase does not appear at any values of $K$ and $L$ in the RSOS5 
model with $E_D=0$. 

\begin{figure}
\centerline{\epsfxsize=6cm \epsfbox{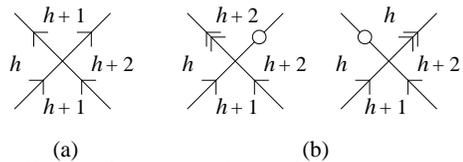}}
\caption{Comparison of configurations of two parallel steps which cost the
interaction energy (a) and from a D step (b). S~(D) steps are
denoted by single~(double) arrows.}\label{fig:3}
\end{figure}
We can argue the reason why the DOF phase disappear in the presence 
of the D step as follows. Consider two parallel S steps 
merging at a vertex. If the D step is not allowed, the possible vertex 
configuration is as shown in Fig.~\ref{fig:3}(a) and the energy cost 
for such configuration is $2K+4L$. On the other hand, 
if the D steps is allowed, the step doubling may occur in two
ways as shown in Fig.~\ref{fig:3}(b) with the energy cost $3K+5L$. 
Though the step doubling costs more energy~($K+L$), entropic 
contribution of the step doubling~($-\ln 2$) may lower the free energy of 
parallel steps below than the value without the step doubling. 
Our numerical results above show that the step doubling suppresses the 
DOF phase entirely in the $E_D=0$ case. In our model a D step costs more
energy than two separate S steps. The two energy scales may be comparable 
to each other in a more realistic model, where the suppression effect will 
be stronger.

\begin{figure}
\centerline{\epsfxsize=70mm \epsfbox{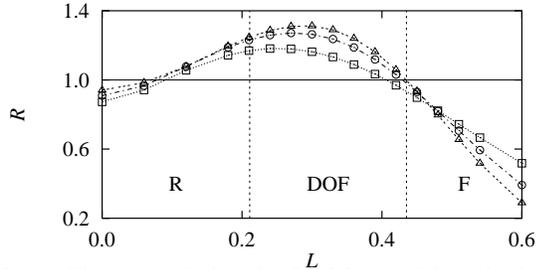}}
\caption{The ratio $R$ for the RSOS5 model with the step doubling energy
$e^{-E_D}=0.2$ along the line $L=5K$. ($\Box$ for $N=4$, $\circ$ for
$N=6$, and $\bigtriangleup$ for $N=8$.)}\label{fig:4}
\end{figure}
From the above arguments, one finds that the step doubling plays an 
important role in phase transitions. So we introduce a new term $E_D N_D$ in
Eq.~(\ref{H_rsos5}) with the step-doubling energy $E_D$ and study the phase
diagram in the parameter space $(K,L,E_D)$. 
When $E_D<0.0$~($>0.0$), the step doubling is
favored~(suppressed).
One can easily expect that the DOF phase does not appear for negative
$E_D$.

For positive $E_D$ the step doubling is suppressed and the effect of the 
step interaction becomes important. So we expect there appears the DOF 
phase in the positive $E_D$ side of the parameter space.
In Fig.~\ref{fig:4} we show the ratio $R$ for $e^{-E_D} = 0.2$ and 
along the line $L=5 K$. Though the convergence is not good, compared with 
Fig.~\ref{fig:2}(a), one can identify three regions as the rough, DOF, 
and flat phases from the $N$ dependence of $R$.
The roughening point between the rough phase and the DOF phase is estimated 
using Eq.~(\ref{fss_pb_r}), and the preroughening point using 
$R=1$ for $N=8$. They are denoted by broken vertical lines.

\begin{figure}
\centerline{\epsfxsize=70mm \epsfbox{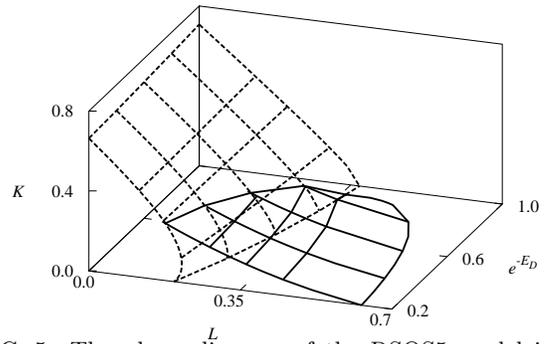}}
\caption{The phase diagram of the RSOS5 model in the $(K,L,e^{-E_D})$
parameter space. The roughening transition lines are denoted by broken lines
and the preroughening transition lines by solid lines.}\label{fig:5}
\end{figure}

We obtain the phase diagram in the whole parameter space 
using the conditions $\eta_{(+,1)} = \frac{\pi \zeta}{4N}$ for the 
roughening transition boundary and $R=1$ for the preroughening transition 
boundary. It is obtained for strip width $N=4,
6$, and $8$. Since the maximum $N$ we can handle is small, the convergence 
of the phase boundary is poor especially as one approaches $e^{-E_D}= 0$. 
But there is no qualitative change in shape. So we only present the phase
diagram obtained from $N=8$ in Fig.~\ref{fig:5}.
The region under the surface represented by broken lines corresponds to the
rough phase. The DOF phase is bounded by the surfaces of broken lines and 
solid lines. The region above the surfaces corresponds to the flat phase.
One should notice that there is a critical value  of $E_D$, approximately
$0.071$, smaller than which the DOF phase does not appear. 

In summary, we have studied the phase transitions in the RSOS5 model
with the Hamiltonian in Eq.~(\ref{H_rsos5}) with D steps as well as 
S steps. We have found that the D
step, which has not been considered in previous works, plays an important role
in phase transitions in crystal surfaces. The presence of the D step 
reduces the strength of the repulsive interaction between parallel steps 
through the step doubling, and hence suppresses the DOF phase. We 
also found that the step-doubling energy is an important 
quantity which characterizes a surface upon the roughening.

I would like to thank D. Kim and M. den Nijs for helpful discussions. 
I wish to acknowledge the financial support of Korea Research 
Foundation made in the program year 1997. This work is also supported by 
the KOSEF through the SRC program of SNU-CTP.

\narrowtext

\end{multicols}
\end{document}